\documentclass[conference]{IEEEtran}
\usepackage{comment}

\ifCLASSINFOpdf
  \usepackage[pdftex]{graphicx}
\else
\fi
\newlength{\codeSpace}
\def\MPIfunclist{
    \begin{list}{}{                     
        \setlength{\leftmargin}{120pt}
        \setlength{\labelwidth}{100pt}
        \setlength{\labelsep}{5pt}
        \setlength{\itemindent}{0pt}
        \setlength{\itemsep}{0pt}
        \setlength{\topsep}{5pt}
    }
}
\def\gb{\penalty10000\hskip 0pt plus 8em\penalty4800\hskip 0pt plus-8em%
\penalty10000}

\newcommand{\funcNoIndex}[1]{\gb\textsf{#1}}
\newcommand{\funcarg}[3]{\item[\hbox to 30pt{\textsf{#1} \hfill} \mpiarg{#2}\hfill]{\small #3}}
\newcommand{\mpiarg}[1]{\gb\textsf{#1}}                  
\newcommand{\IN}[0]{{\small IN}}
\newcommand{\OUT}[0]{{\small OUT}}
\newcommand{\INOUT}[0]{{\small INOUT}}

\newenvironment{funcdef}[1]{
   \vspace{\codeSpace}
    \noindent
    \hangindent 7em\hangafter=1
    {\funcNoIndex{#1}}
    \MPIfunclist
}{\end{list} \vspace{\codeSpace}}

\usepackage{graphicx}
\usepackage{color}
\usepackage[most]{tcolorbox}
\usepackage{tikz}
\usepackage{listings}

\begin{document}
%


\title{Extending the Message Passing Interface (MPI) with User-Level Schedules}

\author{\IEEEauthorblockN{Derek J.\@ Schafer, K.~Sheikh Ghafoor}
\IEEEauthorblockA{Dept.\@ of  Computer Science\\
Tennessee Technological University \\
Cookville, TN, 38505, USA\\
Email: \{djschafer42@students.,\\sghafoor@\}tntech.edu}
\and
\IEEEauthorblockN{Daniel J.\@ Holmes}
\IEEEauthorblockA{EPCC\\
University of Edinburgh\\
Edinburgh, Scotland, UK\\
Email: d.holmes@epcc.ed.ac.uk}
\and
\IEEEauthorblockN{Martin R\"ufenacht, Anthony Skjellum}
\IEEEauthorblockA{SimCenter \& \\ Dept.\@ of Computer Science and Engineering\\
University of Tennessee at Chattanooga\\
Chattanooga, TN, 37403, USA\\
Email: \{martin-ruefenacht,tony-skjellum\}@utc.edu}
}


%


\maketitle

\begin{abstract}
Composability is one of seven reasons for the long-standing and continuing success of MPI. 
Extending MPI by composing its operations with user-level operations provides useful integration with the progress engine and completion notification methods of MPI.  However, the existing extensibility mechanism in MPI (generalized requests) is not widely utilized and has significant drawbacks.

MPI can be generalized via scheduled communication primitives, for example, by utilizing implementation techniques from existing MPI-3 nonblocking collectives and from forthcoming MPI-4 persistent and partitioned APIs. Non-trivial schedules are used internally in some MPI libraries; but, they are not accessible to end-users.

Message-based communication patterns can be built as libraries on top of MPI. Such libraries can have comparable implementation maturity and potentially higher performance than MPI library code, but do not require intimate knowledge of the MPI implementation.  Libraries can provide performance-portable interfaces that cross MPI implementation boundaries. The ability to compose additional user-defined operations using the same progress engine benefits all kinds of general purpose HPC libraries. 

We propose a definition for MPI schedules: a user-level programming model suitable for creating persistent collective communication composed with new application-specific sequences of user-defined operations managed by MPI and fully integrated with MPI progress and completion notification. The API proposed offers a path to standardization for extensible communication schedules involving user-defined operations. Our approach has the potential to introduce event-driven programming into MPI (beyond the tools interface), although connecting schedules with events comprises future work.  

Early performance results described here are promising and indicate strong overlap potential.

\end{abstract}

\IEEEpeerreviewmaketitle

\section{Introduction}
\label{sec:introduction}
It is useful to extend MPI communication operations with new and composed operations that integrate with the progress engine and completion notification methods of MPI. But,
generalized requests, defined in MPI-2 to support such extensions,
are not fit-for-purpose
and are rarely
used in practice.
The advent of persistent collective communication and forthcoming partitioned point-to-point communication opens an opportunity for a new,
high performance
solution.

The key contributions of this paper are as follows: MPI can  be generalized by the introduction 
of scheduled communication primitives that utilize forthcoming MPI-4 persistent APIs.
Non-trivial schedules 
are already used internally in some MPI libraries, but they are not accessible to end users as first-class programming constructs. These constructs will prove most useful when coupled with a strong progress engine and blocking completion notification. In many cases, MPI library writers will be able to add message-based collectives of comparable implementation maturity and potentially higher performance to those implemented internally in an MPI library, yet without intimate knowledge of the implementation or need to augment it internally. Those algorithms would provide a new kind of performance-portable MPI code that crosses MPI implementation boundaries. MPI 
schedules provide a user-level programming model suitable for creating persistent collective communication, nonblocking communication, plus new, application-specific sequences of operations in MPI. While schedules 
exist within some implementations of MPI, our approach provides a user-level API that integrates with MPI progress and completion notification. The API proposed offers a path to standardization for 
extensible communication sequences
with the potential to introduce event-driven programming as a future extension to MPI (not just tools-related). Means for connecting scheduled communication with internal and external events is considered for future work.

The remainder of this paper is organized as follows: Section~\ref{sec:background} describes the motivations and background for this work.  Section~\ref{sec:design} describes the design of user-level schedule communication including API specifications. Section~\ref{sec:modeling} describes qualitative performance modeling.  Implementation issues and details, including how this work builds on the
ExaMPI~\cite{CARLA2019}
research implementation of MPI, are described in Section~\ref{sec:implementation}. Section~\ref{sec:results} presents some preliminary results with user-level schedules. Section~\ref{sec:future} describes possible extensions to user-level scheduled communication and indicates plans for future work, while Section~\ref{sec:conclusion} offers conclusions.   

\section{Background and Motivation}

\label{sec:background}
This section addresses background concepts and motivations for this paper, including how this fits with forthcoming additions to MPI, and where it builds on previous design, implementation, and/or standardization work.

\subsection{Motivations}
The following properties of MPI-4 (accepted and/or proposed functionality), inform this effort:
\begin{itemize}
\item Collective communication operations approved for MPI have initialization operations. Such INITs are currently collective (non-local, may synchronize) because of effects involving hand-off between the application and MPI of array parameters (e.g., displacements) to certain collectives plus the non-blocking properties of MPI\_REQUEST\_FREE.
The future addition of local variants of these operations will not break backward compatibility, but will require a stronger completion form of MPI\_REQUEST\_FREE to ensure that resources held by MPI are relinquished to the user.
Also, addition of local variants of currently non-local functionality must be considered
throughout the MPI Standard
(e.g., MPI\_WIN\_ICREATE, etc). The present semantic compromise is clearly shown in \cite{TsAndCs2019}%
;  the Venn diagram(s) depicted there that show collective initialization procedures as incomplete and non-local whereas collective initiation procedures and point-to-point initialization procedures are both shown as \emph{incomplete} and \emph{local}.

\item Channels---point-to-point communication between two MPI processes---will be achieved in MPI via partitioned communication \cite{DBLP:conf/supercomputer/GrantDLBS19}, which is presently being considered for MPI-4. Partitioned communication, in the fullness of time, will have blocking, nonblocking, and persistent API variants for point-to-point, collective, one-sided, and I/O operations. Persistent partitioned communication adds a new concern to planning schedules (the key goal of this paper):
the user may inform MPI that individual parts of an outgoing message are ready for transmission without being required to create and manage multiple messages. A similar semantic extension, the ability to consume individual parts of an incoming message without being required to create and manage multiple messages, is also being considered.

Breaking an MPI operation into pieces that are scheduled separately is not novel. However, allowing the user to
indicate
when some of the pieces are ready, as proposed for partitioned communication, is new and it presents a new challenge for planning schedules. Naively, partitioning could be scheduled in the same manner as separate operations, that is, one operation per partition.
In that case, the completion of the whole partitioned operation is achieved when all the separate per-partition operations have completed - much like a call to MPI\_WAITALL but without all the MPI\_REQUEST objects.
Benchmarking work for partitioned communication~\cite{DBLP:conf/supercomputer/GrantDLBS19} shows that the separate messages approach is less efficient than a straightforward implementation of the partitioned approach.
Less naively, user-driven execution of parts of a schedule comprises a DAG schedule with multiple root vertices (one per partition), where each root vertex requires a user down-call to satisfy its input dependencies. This seems to present a major challenge to the linearization-into-command-queue technique used, for example, in LibNBC  \cite{libNBC} and in Open MPI \cite{gabriel04:_open_mpi}.
Any technique to store an execution plan or schedule for MPI operations is likely to have to take this new consideration into account - requiring structural changes that acknowledge the new sources of input dependencies between the scheduled sub-tasks.

\item Some communication patterns involve multiple persistent requests and managing these requests could become cumbersome. For example, a user might want to perform entire communication protocols that are not defined by MPI (e.g., halo exchange for domain decomposition) with an all-reduce (e.g., for monitoring conservation of energy/mass) plus another all-reduce (e.g., for termination criteria) before finishing off with collective I/O write (such asfor checkpointing). Or, a user might simply want to do a ``gather-scatter'' communication pattern as done in~\cite{Gong2016}. Either way, such patterns would benefit from having a means to combine multiple persistent requests into a single, manageable persistent request. With schedules, a user could easily define new collectives not defined by MPI as well as provide new implementations of existing collective operation that are defined by MPI.



We recognize three options for API design here: 
\begin{enumerate} 
\item assert
 that generalized requests, as currently defined, solves this need already (we discuss why this is not ideal in Subsection~\ref{subsec:grequests} below); 
\item allow the user to associate existing persistent requests together with each other somehow (e.g., with INFO to each initialization procedure warning MPI that this request will be cobbled together with others), plus a new set of mechanisms 
to achieve that connectivity; 
\item design and expose a new API that provides to the user the schedule creation and manipulation functionality that is currently internal-only inside  MPI libraries (e.g., expose all the internal operations in Open MPI \cite{gabriel04:_open_mpi} (which derived from LibNBC \cite{libNBC}): ompi\_sched\_add\_send, … functions, renamed with MPI prefix, of course, and add new functions to manage the lifetime of MPIX\_Schedule objects, like:
 \begin{itemize} \item MPIX\_Schedule\_create(inout SCHED)
 \item MPIX\_Schedule\_commit(in SCHED, \\  \hspace*{1.4in} out MPI\_Request)
 \item MPIX\_Schedule\_free(inout SCHED)
 \end{itemize}
 \end{enumerate}
In this paper, we employ a hybrid of the latter two approaches, while discarding the current formulation of generalized requests as a viable option.
\end{itemize}
The next subsections discuss other MPI technologies and further background and motivations for this work.
\subsection{MPI Generalized Requests}
\label{subsec:grequests}

MPI provides baseline support for generalized requests \cite{MPI-3.1}. An MPI generalized request provides users with the ability to develop their own nonblocking functions for both point-to-point and collective scenarios. However, the MPI documentation states that the OS is responsible for concurrent execution, and thus MPI only supports ways of enabling concurrency mechanisms to interact with MPI. As such, generalized requests are defined by a set of user-provided functions that are called when certain MPI actions (i.e., MPI\_WAIT, MPI\_REQUEST\_FREE, MPI\_CANCEL, etc.) are performed on the generalized request. An example use-case of generalized requests in MPI is when an application has both a communication thread and a computation thread. While the application is responsible for making and managing these threads, the communication thread has the same means to interact with MPI as the computation thread would. The computation thread can call certain MPI functions to check on the progress of the request, but cannot be notified by MPI about any intermittent progress of the request. In addition to managing the communication, the communication thread would also be responsible for spending time progressing the generalized request and marking it as complete later, since MPI does not allow the user to hook their generalized request into the progress engine in any meaningful way. Furthermore, the additional thread progressing the generalized request is also responsible for performing the operations of the request (i.e., making the point-to-point, collective MPI calls, local operations, etc.).
Thus, the drawback to generalized requests can be summed up with how the interface provided by MPI requires that the user make independent progress for all generalized requests and then inform MPI when each one is complete. 
Certain efforts to generalize generalized requests has also been done  (e.g., \cite{DBLP:conf/pvm/LathamGRT07}).

We note that, of the 110 the MPI applications studied by Laguna et al.\@ in \cite{Laguna2019}, none
of the  applications analyzed 
utilize MPI generalized requests.

\subsection{LibNBC and Open MPI Schedules}
Hoefler and Lumsdaine created LibNBC~\cite{libNBC} with the goal of creating non-blocking collective functions that enabled a better overlap of communication and computation. To build a collective within LibNBC, a series of operations and rounds must be combined to form the new collective's schedule. The rounds are ordered sequentially and all operations in a round must be completed before the round can move on to the next. A collective implementation within the library calls API functions to add different operations to the current round, switch to the next round, and to commit the schedule before starting it. Collective operations created within LibNBC are done in a separate context of the communicator to avoid interfering with regular user communication in MPI. Progress inside a round can be asynchronous. However, in order to progress to the next round, one of LibNBC's testing functions must be called, limiting its ability to be truly asynchronous.

LibPNBC~\cite{DBLP:conf/pvm/MorganHSBS17,HOLMES201932}, developed by Morgan et al., is an extension to LibNBC that provides support for persistent collective operations in MPI. Open MPI derives its scheduling mechanism from LibNBC, which it also incorporates and generalizes as a component of that MPI implementation~\cite{gabriel04:_open_mpi}.  Users  cannot access LibNBC to create schedules; the LibNBC API is not published at the user-level by Open MPI; it is used internally.

The user-level schedules proposed in this paper are better than libNBC/libPNBC/OMPI-schedules because they are user-accessible, support run-once starting and ending operations, and work with persistent operations, rather than late-binding MPI operations. (In future work, we we will make them fully Turing complete.)


\section{Design}
\label{sec:design}

Our goal in providing 
extensions and enhancements for 
scheduled operations is to support both on-loaded and off\-loaded operations.
If one can offload a large fraction of a schedule for an operation to switches, for instance, then terabit line rates (with commensurate message-rates) become possible for MPI.
Removing the CPU from the critical path as much as possible is a strong goal moving forward, especially for multi-stage communication operations that can be planned with immutable schedules.
This paper, without loss of generality, focuses on on-loaded schedules for communication operations (executed in software by the CPU).

In the remainder of this section, we cover the syntax and semantics of our proposed extensions to MPI for user-level scheduled communication.
\subsection {Operation APIs - Syntax and Semantics}

The design of the API is similar to that of LibNBC, and follows similar semantics with rounds and operations, at least at the user level (although, as an important distinction, our operations are always persistent MPI-4 operations\footnote{Relaxing this restriction is left for future work.}).  Users can create a round, add as many operations to it as necessary, and then repeat the process to make as many rounds as needed. Once the user is satisfied with the schedule, they can commit the schedule to finalize it and get a schedule request (which is a persistent request). Once the user is ready to launch the operation, the user can start  request using MPI\_START (or MPI\_STARTALL, if appropriate). One notable difference from LibNBC is that the sub-requests introduced in the schedule do not have to progressed by user code; instead, they are progressed by MPI's progress engine. Another distinction is how the requests are added to the schedule. This API requires users first to  initialize all of the requests they wish to use in the schedule (as opposed to having specific API functions for adding each operation, like NBC\_Sched\_send). Through the use of functions like MPI\_BCAST\_INIT (presented in the MPI-4.0 standard), current persistent operations, like MPI\_SEND\_INIT, and partitioned communication functions, like MPIX\_PARTITIONED\_SEND\_INIT from finepoints~\cite{DBLP:conf/supercomputer/GrantDLBS19}, users obtain a request handle to the specific operation, and then can pass that request handle to the schedule to insert it as a sub-request. The remainder of this section  explores the functions that are proposed for the user-level API.

\begin{funcdef}{MPIX\_SCHEDULE\_CREATE(schedule, auto\_free)}
\funcarg{\INOUT}{schedule}{schedule handle}
\funcarg{\IN}{auto\_free}{a boolean indicating auto-freeing}
\end{funcdef}

MPIX\_SCHEDULE\_CREATE
is the first function that a user would use to build a schedule. This function is designed to create an empty schedule and return a handle  so they can begin using rounds and operations to build their operations incrementally. 
The created schedule also has the option to automatically free its requests when the schedule itself is freed.



\begin{funcdef}{MPIX\_SCHEDULE\_ADD\_OPERATION(schedule, request, auto\_free)}
\funcarg{\INOUT}{schedule}{schedule handle}
\funcarg{\IN}{request}{request handle}
\funcarg{\IN}{auto\_free}{a boolean indicating auto-freeing}
\end{funcdef}

\vspace*{-.1cm}
Once the user has obtained a request for some operation that they wish to add to the schedule, the request can be passed to MPIX\_SCHEDULE\_ADD\_OPERATION to be added as a sub-request to the specified schedule. 
Request handles returned from MPI\_SEND\_INIT and other communications operations are an example of operations that can be added this way.

%
Like the schedule itself, one can mark the requests as auto-freeing, which means that when the schedule request is freed, any sub-request that is marked auto free will also be freed. Any request that is not marked will allow the user to regain use of a valid, inactive handle after freeing the schedule. More specific semantics are discussed in the next section.


\begin{funcdef}{MPIX\_SCHEDULE\_ADD\_MPI\_OPERATION(schedule, mpi\_op, invec, inoutvec, len, datatype)}
\funcarg{\INOUT}{schedule}{schedule handle}
\funcarg{\IN}{mpi\_op}{the MPI\_Op to perform}
\funcarg{\IN}{invec}{the first operand}
\funcarg{\INOUT}{inoutvec}{the second operand}
\funcarg{\IN}{len}{how many items to perform operation on}
\funcarg{\IN}{datatype}{the datatype of the operands}
\end{funcdef}

Additionally, a user may wish to perform some sort of operation on the data between communication steps. In fact, in order to properly build a reduction into a schedule, a user must have some way to perform a calculation on the data between steps. This function provides that capability by supporting the addition of any MPI\_Op to the schedule. The user specifies the MPI\_Op (such as MPI\_MAX, MPI\_SUM, etc.) and the location of the two operands. The results of the operation are stored at the location of inoutvec. User can also provide any user-created MPI\_Op to the schedule\footnote{For the specific semantics of user-defined operations, see Section 5.9.5 in the MPI 3.1 standard~\cite{MPI-3.1}.}. In this case, the progress engine will call the user function described in the operation and pass the last four parameters to the user function.

\begin{funcdef}{MPIX\_SCHEDULE\_MARK\_RESET\_POINT(\linebreak{}schedule)}
\funcarg{\INOUT}{schedule}{schedule handle}
\end{funcdef}

MPIX\_SCHEDULE\_MARK\_RESET\_POINT allows the user to specify that all rounds before the current round constitute rounds with operations that are only ever needed to be done the first time the schedule is executed\footnote{
This concept does not exist in LibNBC/Open MPI schedules.}. Such rounds could be used to perform setup or synchronization before the main workload is to be performed.
Every time the schedule is launched afterwards, the schedule will start from the round after the reset point. Calling this function several times while building a schedule will simply move the reset point to the current round. If a user never calls this function when creating their schedule, the reset point is by default the first round; all rounds will be performed every time the schedule is executed.

\begin{funcdef}{MPIX\_SCHEDULE\_MARK\_COMPLETION\_POINT(\linebreak{}schedule)}
\funcarg{\INOUT}{schedule}{schedule handle}
\end{funcdef}

Completion points allow the user to specify that any further rounds will only be executed the final time a given schedule is executed. When a schedule reaches this point, it will be treated as the end of the schedule and be reset back to the reset point. The execute-once portion of the schedule that is after the completion point will only be executed by the progress engine once the request has been marked for freeing using the MPI\_REQUEST\_FREE function or during MPI\_FINALIZE (whichever occurs first). There is currently no way in MPI to guarantee resource recovery before MPI\_FINALIZE. However, it is good practice to call the relevant freeing function as soon as the object or handle is no longer needed, as it gives MPI the opportunity to execute clean-up actions earlier. Similar rules as MPIX\_SCHEDULE\_MARK\_RESET\_POINT apply when trying to mark multiple completion points. If the function is never used on a schedule, then the last round is assumed to be the completion point. One could use a round after the completion point to perform tear-down or synchronizing action.

\begin{funcdef}{MPIX\_SCHEDULE\_CREATE\_ROUND(schedule)}
\funcarg{\INOUT}{schedule}{schedule handle}
\end{funcdef}

After adding several operations to the schedule, the user may wish to move onto the next round. To do so, the user would call MPIX\_SCHEDULE\_CREATE\_ROUND. This function ends the current round by adding a new round to the schedule. Any operations that are added after this call are put into the new round. It is not valid to have an empty round in the schedule, so calling MPIX\_SCHEDULE\_CREATE\_ROUND when the current round is empty will not create an additional round.  Once the user is finished with the overall schedule, they do not need to call MPIX\_SCHEDULE\_CREATE\_ROUND an additional time to ``finalize'' the last round. They should instead call MPIX\_SCHEDULE\_COMMIT.

\vspace*{-.1cm}
\begin{funcdef}{MPIX\_SCHEDULE\_COMMIT(schedule, request)}
\funcarg{\IN}{schedule}{schedule handle}
\funcarg{\OUT}{request}{request handle}
\end{funcdef}

MPIX\_SCHEDULE\_COMMIT should be called when the user has completed the building phase. When the user calls this function, they receive a handle to a request that can be passed to MPI\_START to launch the schedule. Additionally, this request can be passed to MPI completing functions, such as MPI\_WAIT and MPI\_TEST. A user could also potentially nest schedules together to create a hierarchy of operations by passing this reuqest into another schedule. Should a schedule be committed with the current round lacking any operations to perform, the empty round will be trimmed from the schedule since there is nothing for the progress engine to do. If commit is called on a  schedule with only one round and no operations, an error will be returned and the request handle will not be valid. After calling MPIX\_SCHEDULE\_COMMIT, it is not valid to add more operations, rounds, reset points, or completion points to the schedule.

\begin{funcdef}{MPIX\_SCHEDULE\_FREE(schedule)}
\funcarg{\INOUT}{schedule}{schedule handle}
\end{funcdef}

\vspace*{-.1cm}
This function is used to free the schedule specified. After a schedule has been freed, it is not valid to perform any more actions on the schedule. Any corresponding request created using the freed schedule can be also freed with MPI\_REQUEST\_FREE. Additional considerations for freeing the requests will be discussed in the next section.

%

\subsection{Operational Semantics}
The prior section presents an overview of the API functions we  propose, along with description of these operations and constraints on their use. This section discusses  additional considerations about the properties of the proposed 
API that affect its operational semantics:
\vspace*{-.1cm}
\begin{enumerate}
    \item 
Once a request is bound to a schedule (and thereby used in a new composite communication operation), it can no longer be used outside of that schedule. Only one schedule can own a request at a time; it is not valid to add the same request to two different schedules. However, because our API requires the user to provide every request they wish to add to the schedule, the user will have access to the request handles after adding them to the schedule. As these requests are normal MPI\_Requests, allowing the users to call some MPI functions on them could prove to be useful. It is reasonable that a user might find it useful to check on the progress of a specific sub-request inside the schedule. Thus, we have 
required that operations such as MPI\_WAIT and MPI\_TEST remain valid on requests that are a part of the schedule. This allows a user to retrain fine-grained control over the application, by allowing processes to react to the completion of specific parts of the schedule.
However, calling MPI functions that could result in the individual requests being modified, such as MPI\_REQUEST\_FREE and MPI\_START, is not permitted. Calling these functions on requests currently within a schedule could break the concurrency provided by that schedule or produce undefined behavior on the resources associated with the requests.
    \item
The steps of creating, building, and committing a schedule together constitute the initialization stage of a persistent MPI operation. The resulting 
 MPI\_Request
handle represents that persistent operation, which can be used to start the operation (e.g., using  
MPI\_START)
to complete the operation (e.g., using either  
MPI\_TEST
 or  
MPI\_WAIT),
and to free the operation (i.e., using  
MPI\_REQUEST\_FREE).
The rules for starting a persistent operation created by committing a schedule depend on the rules for starting the constituent operations from which it was built.
If any of the constituent operations requires a particular start order, then the composite operation must also adhere to that restriction.
For example, if a persistent point-to-point operation is added to a schedule, then the resulting composite operation must be started at a time (and in an order relative to other MPI function calls) that permits correct matching of the point-to-point operation with the corresponding operation at the destination (or source) MPI process.
Similarly, if a persistent collective operation is added to a schedule, then the resulting composite operation must be started at a time (and in an order relative to other MPI function calls) that complies with the value given for the 
mpi\_assert\_strict\_start\_order
MPI\_Info
key.

If the mpi\_assert\_strict\_start\_order assertion is not used with a given  persistent collective operation (that is, defined on the communicator from which the operation was created), then MPI has complete freedom of ordering with regard to all other operations MPI does with respect to each process involved.  If that assertion is used, however, then the scheduled operation must be started across the group of each communicator associated with it in a way to keep the start order consistent across the whole group.
    \item 
When building the schedule and the request objects, it makes sense that, at some point, these objects will need to be freed. As touched on above, both the schedule and the request have a parameter that allow the user to specify whether or not they will be freed when a function is called. By default, all requests will be automatically freed when the schedule request is freed. As such, the sub-requests will not be viable  after the conglomerate operation is freed\footnote{Freeing described is in respect to how the MPI Standard talks about MPI managing resources from the user. While some languages have the ability to handle allocation and de-allocation of resources for the user, MPI itself generally does not. Thus, to maintain portability, we provide these freeing options}.
\end{enumerate}

\begin{figure}[t]
    \centering
    \includegraphics[width=.85\linewidth]{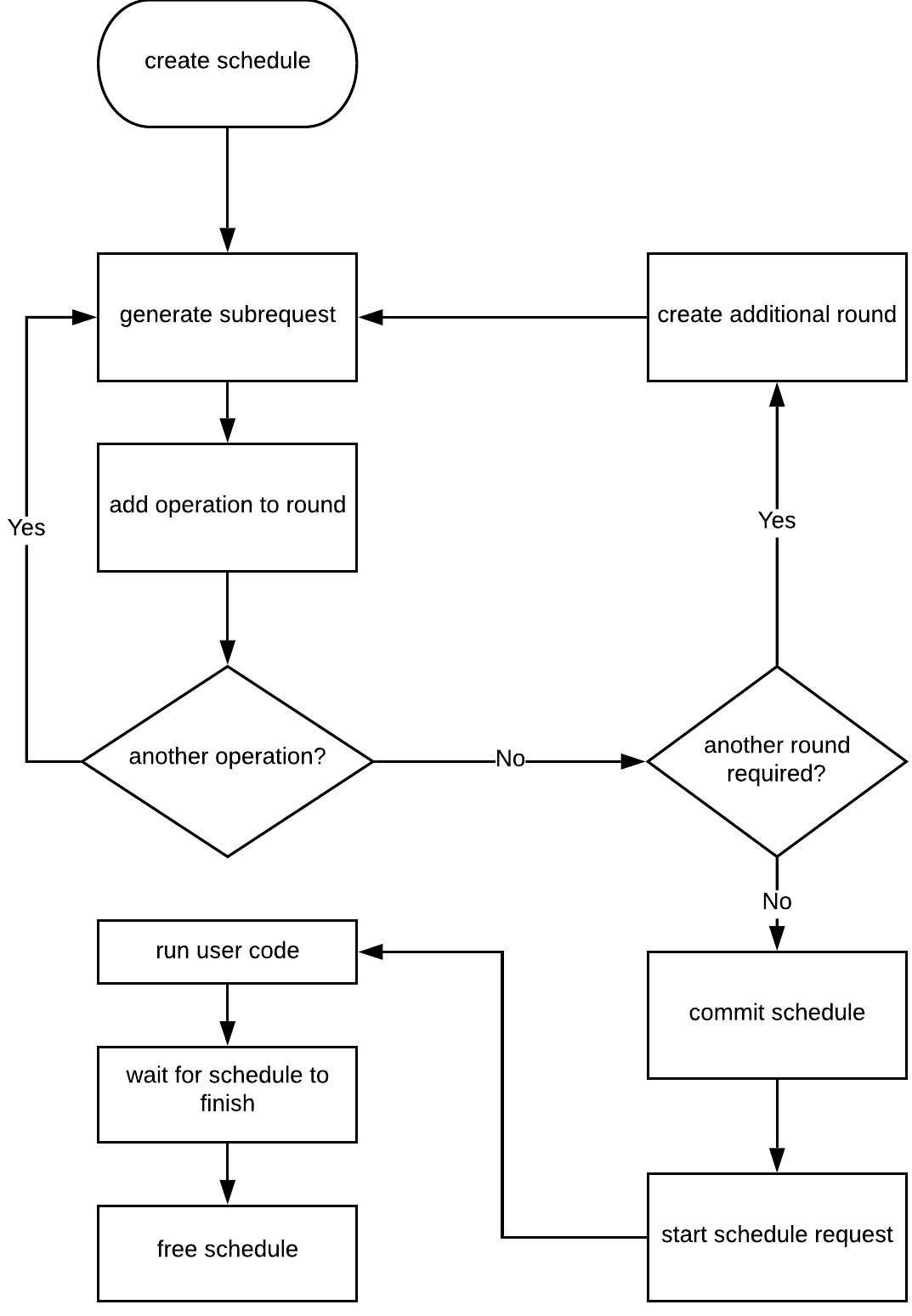}
    \caption{Procedure of schedule creation}
    \label{fig:schedule}
\end{figure}

\vspace*{-.1cm}
\subsection{Examples}

An example flow for building a schedule is shown in Fig.~\ref{fig:schedule}. Using this flow,  sample pseudo-code is presented in Fig.~\ref{fig:code_example}. This code shows how to combine the new API with the existing concept of adding sends, receives, and collectives to the schedule from LibNBC. Roughly, this code is building a schedule for a reduction to non-zero rank from a reduction to zero-rank and a point-to-point operation. All parameters not directly related to building the schedule
are elided
for brevity. The use of auto-freeing has been omitted for brevity.

\begin{figure}[t]
\definecolor{mygreen}{rgb}{0,0.6,0}

\lstset{ %
  backgroundcolor=\color{white},   
  basicstyle=\footnotesize,        
  breaklines=true,                 
  captionpos=b,                    
  commentstyle=\color{mygreen},    
  keywordstyle=\color{blue},       
  stringstyle=\color{red},     
}

\begin{lstlisting}[language=C++]
//Create & initialize schedule
MPIX_Schedule sched;
MPIX_Schedule_init(&sched);

//Add reduce operation to schedule
MPI_Request reduce_op;
MPI_Reduce_init(<reduce parameters, root = 0>, &reduce_op);
MPIX_Schedule_add_operation(&sched, reduce_op);

//Prior ops only happen on first run
MPIX_Schedule_mark_reset_point(&sched);

MPIX_Schedule_create_round(&sched);

//Add more operations
if (0 == my_rank) {
    MPI_Request ssr;
    MPI_Send_init(<send parameters w/ dest = root>, &ssr);
    MPIX_Schedule_add_operation(&sched, &ssr);
}
else if (root_rank == my_rank) {
    MPI_Request rsr;
    MPI_Recv_init(<recv parameters - source = 0>, &rsr);
    MPIX_Schedule_add_operation(&sched, &rsr);
}
//Create request handle for schedule
MPI_Request request;
MPIX_Schedule_commit(&sched, &request);

//Start schedule
MPI_Start(request);

//   Other user code to run   //
// while schedule is executed //

//Wait for schedule to complete
MPI_Wait(request, &status);
MPI_Request_free(&request);
MPIX_Schedule_free(&sched);

\end{lstlisting}
\caption{Scheduled communication example}
\label{fig:code_example}
\end{figure}

\vspace*{-.1cm}
\section{Qualitatively Modeling Performance}
\label{sec:modeling}

\begin{figure*}[t]
    \centering
    \includegraphics[width=\linewidth]{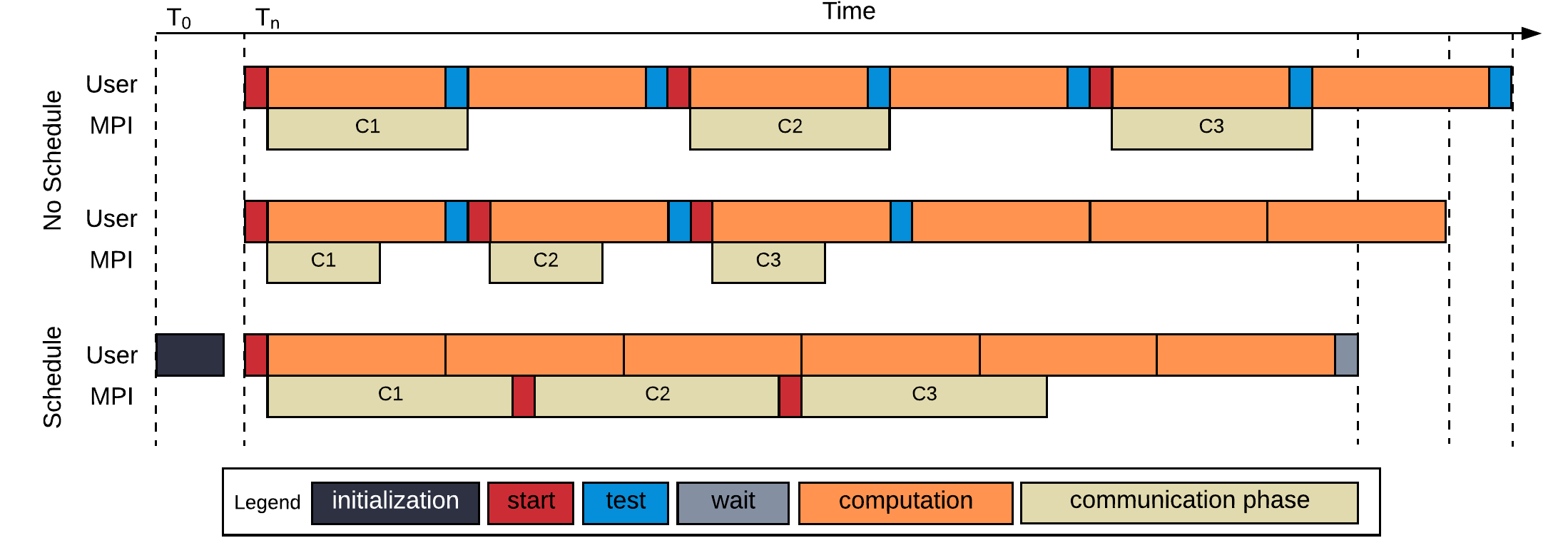}
    \caption{A comparison between a user manually progressing communication and a user using a schedule (bottom). All computational blocks are of equal work. Both the worst-case for manual progression, where communication ends just after the user checks (top), and the best-case, where communication ends before the user checks (bottom), are presented. As shown, the use of a user-level schedule allows for a reduced time-to-solution due to the removal of test operations. A one-time cost of building the schedule is now required, but is inexpensive if the schedule is used multiple times.}
    \label{fig:pipe1}
\end{figure*}

In this section, we consider qualitative modeling concepts to show the value of user-level scheduled requests. If those scheduled requests are offloaded to a progress engine that works asynchronously from the user thread, a considerable amount of time can be saved, especially if the schedule is persistent and used often. Fig.~\ref{fig:pipe1} shows 
how this is possible by demonstrating the timing of one application cycle. Note that after the first cycle, all applications will start from the same spot, time Tn.

In this example, a user application has six equally sized workloads to do, all of which can be done independently from any communication needed to be done. At the same time, the user application also needs to do three communication sequences. While not specifically intertwined, these nine operations must be completed before the user application can proceed to the next step of the code, and these nine operations will be repeated many times during the user's application. In current MPI code, an application wishing to start this process would first start the first communication sequence and then begin computations. In the first example in Fig.~\ref{fig:pipe1}, the user's application completes the first computation sequence just before the first communication sequence finishes. When the user tests to see if the communication sequence has completed, they see that it has not and proceed to start another computation sequence. After completing those computations, the user application tests again, finds that the communication sequence has finished, and starts the next round of communication. Unfortunately for the user, the communication sequences continue to run long, and the application achieves the worst case scenario of communication and computation overlap.

The second example in the figure represents the opposite scenario; now the user application has achieved   good overlap between communication and computation as a result of communication times being quicker. Here, every time the user is ready to test for completion of a communication sequence, they find that it has completed and can immediately progress communication to the next sequence. Additionally the user is also able to finish all necessary communications after its third computation sequence, and thus does not have to spend any extra time by testing and launching more communications later on (at least in the context of this portion of the code).  This example shows how achieving  good communication and computation overlap can help the program achieve a shorter time to completion. However, we can do even better with user-level scheduled requests.

Consider the third example in the diagram. Here the user application is using an MPI implementation
that supports user-level schedules. In this code, 
the user absorbs the one-time cost spent on creating the persistent communication sequences (building the schedule, time T0 to Tn) but, 
does not have to spend any time subsequently testing for communication completion. Once created, the user must also offload this communication sequence to the progress engine. 
Afterwards, the user is free to do all the computation sequences, and does not need to waste time progressing the communication sequences, as the progress engine uses the provided schedule to do so. When the application is ready to check on the progress, it can wait to see if the schedule is complete. 
Since the schedule is persistent, the user can use the schedule right away on subsequent uses; there is no cost to restart the schedule, only the initialization and building cost on the first use.


\begin{figure*}[t]
    \centering
    \includegraphics[width=\linewidth]{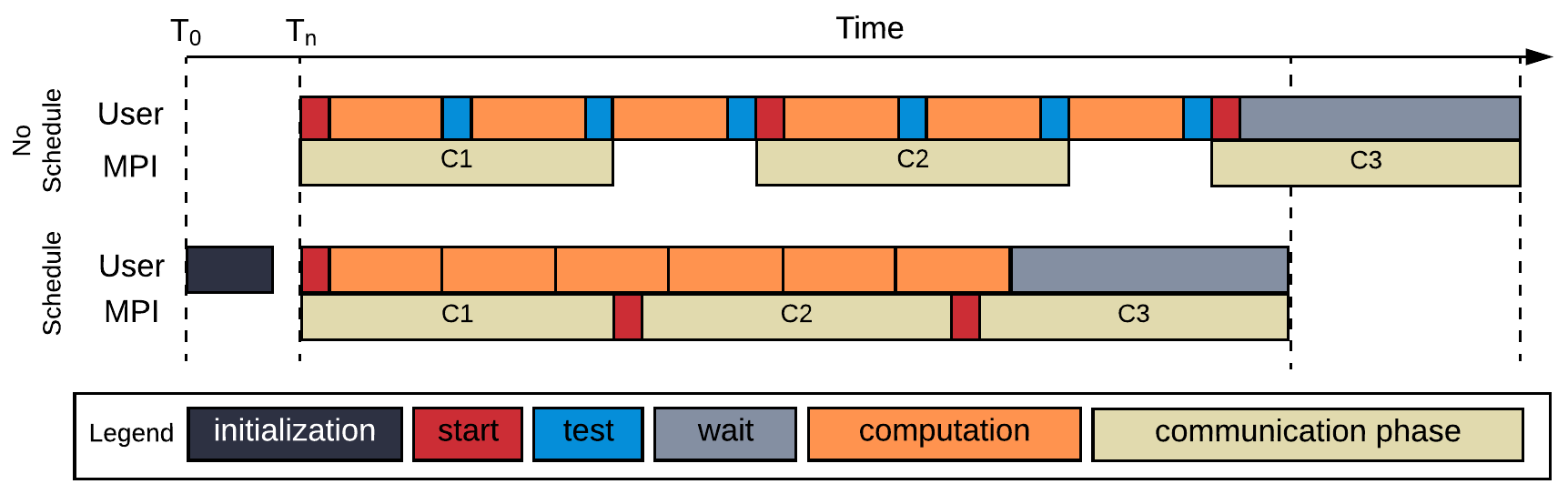}
     \caption{Another comparison between a user manually progressing communication (top) and a user using a schedule (bottom). All computational and communication blocks are of equal size. As shown, offloaded schedules allow for reduced time-to-solution due to the removal of test operations by the user and the ability of the user to leverage the progress engine to progress communication. The overhead of initialization is again amortized across repetitions of the offloaded schedule.}
    \label{fig:pipe2}
\end{figure*}

The above examples assume that the computation takes longer than the communication. Next, we illustrate that even with longer communication time than computation time, better overlap is still achieved by using the schedule alongside a strong progress engine. Fig.~\ref{fig:pipe2} depicts this scenario. In this situation, both scenarios end up completing all possible computation they can, and then must wait on the communication sequences to complete.
However, when the application is directly managing all the constituent communication operations, it must spend time manually progressing multiple requests (through repeated calls into MPI completing procedures, like MPI\_TEST).
For the program using a schedule, the progress engine takes care of moving all the constituent communication sequences along.

\section{Implementation}
\label{sec:implementation}
In this section, we first describe the underlying MPI implementation, ExaMPI. We focus on its key features, as well as its strong progress engine and support for schedules. With that understanding, we then discuss the implementation of schedules in ExaMPI.

\subsection{ExaMPI}
\begin{figure}[t]
    \centering
    \includegraphics[width=\linewidth, trim=0 0.45cm 0 0.5cm, clip]{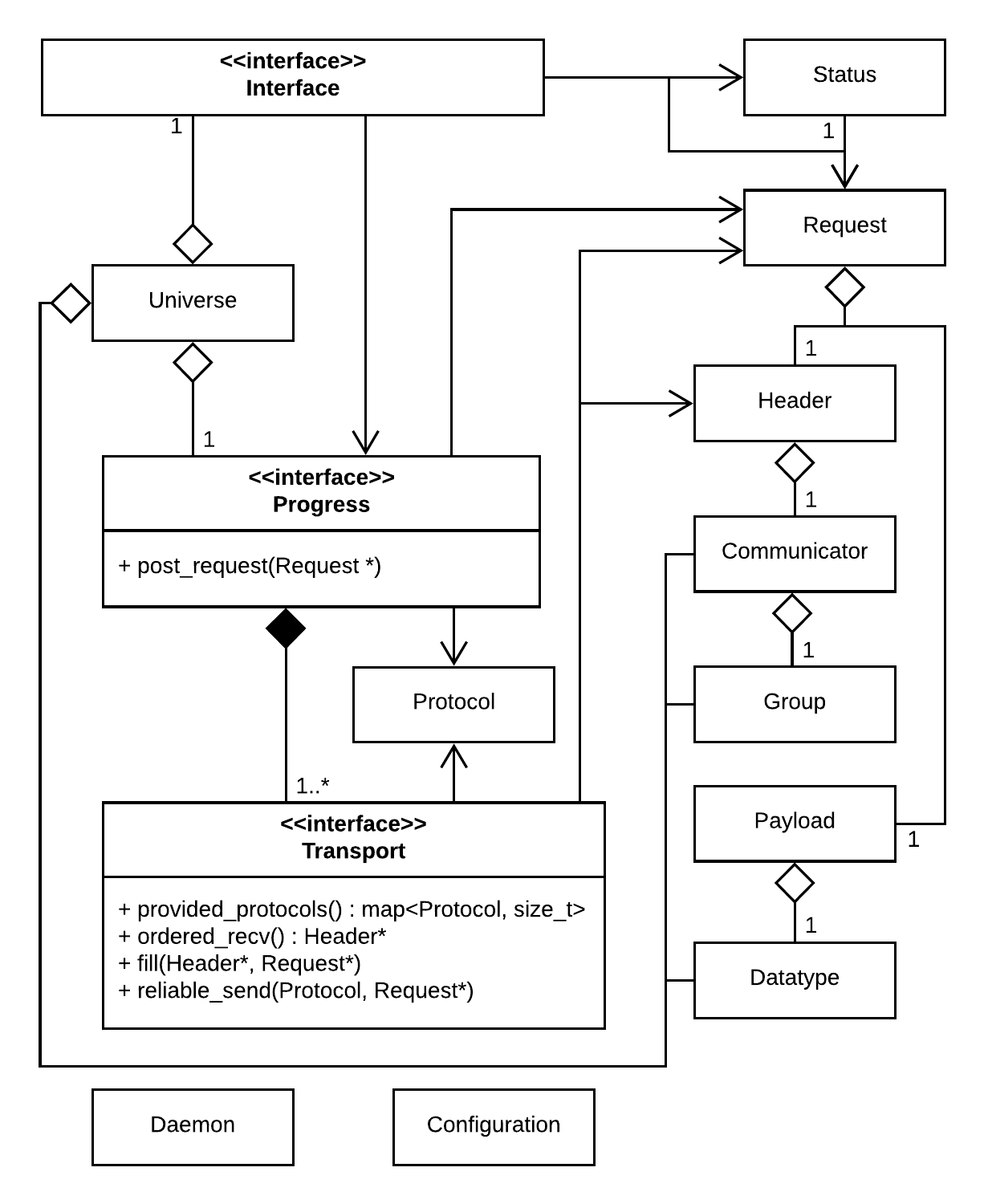} 
    \caption{Design overview of ExaMPI's structure (adopted from \cite{CARLA2019})}
    \label{fig:exampi-highlevel-uml}
\end{figure}

\definecolor{same_green}{RGB}{5,142,217}

\begin{figure}[t]
    \centering
    \begin{tikzpicture}[scale=.675] 
        \node at (4,8.8) {\large \bf Progress};
        \node at (2,8.3) {\large Independent};
        \node at (6,8.3) {\large Polling};
        
        \node[rotate=-90] at (9,4) {\large \bf Notification};
        \node[rotate=-90] at (8.3,6) {\large Blocking};
        \node[rotate=-90] at (8.3,2) {\large Polling};
    
        \filldraw[fill=same_green, draw=black] (0,0) rectangle (4,4);
        \filldraw[fill=same_green, draw=black] (4,0) rectangle (8,4);
        \filldraw[fill=same_green, draw=black] (0,4) rectangle (4,8);
        \filldraw[fill=white, draw=black] (4,4) rectangle (8,8);
        
        \node[align=center] at (6,2) {weak progress \\ (MPICH, \\ OpenMPI, \\MPI/Pro, \\ ExaMPI*)}; 
        \node[align=center] at (2,6) {strong progress \\ (ExaMPI,\\ MPI/Pro)}; 
        \node[align=center] at (2,2) {saturated progress \\ (MPI/Pro, \\ ExaMPI*)}; 
        \node[align=center] at (6,6) {anti-progress}; 
    \end{tikzpicture}
    \caption{Dimitrov's Progress and Notification Classification Diagram (adopted from \cite{Dimitrov2001,CARLA2019}); *Forthcoming modes in ExaMPI in Fall 2019.} 
    \label{fig:dimitrovclass}
\end{figure}
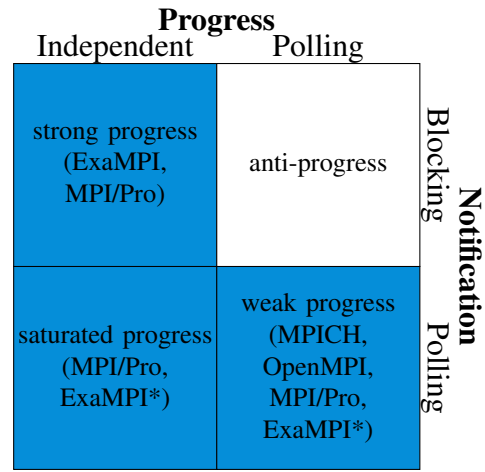

ExaMPI \cite{CARLA2019} is a C++17-based library  designed for modularity, extensibility, and understandability. ExaMPI's high-level structure is depicted in Fig.~\ref{fig:exampi-highlevel-uml}. The design ensures internal interfaces through C++ classes, which allows replacement of behavior with ease. The code base supports both native C++ threading with thread-safe data structures and a modular progress engine. In addition, the transport abstraction implements UDP, TCP, OFED verbs, and LibFabrics for high-performance networks.

The progress engine interface shown in Fig.~\ref{fig:exampi-highlevel-uml} specifies a minimal set of functions required. Through this interface, all modes of progress and completion are supported as depicted in Fig.~\ref{fig:dimitrovclass}. The strong progress engine present in ExaMPI allows a variable number of progress threads and supports more modularity internally to support different matching algorithms and decision functions about operation algorithm usage. The overall design allows for offloading of any operations or even the entire progress engine onto hardware.

Despite being an immature implementation that is not perfectly tuned yet, we still chose to implement user-level schedules in ExaMPI due its support for a strong progress engine. Without a strong progress engine, we cannot experiment with the offloading potential of the schedules and achieve the benefits outlined in Fig.~\ref{fig:pipe1} and Fig.~\ref{fig:pipe2}.

\subsection{Building New Operations}
As previously discussed, a schedule consists of several rounds, each containing at least one operation. A round contains one or more sub-requests that can be launched concurrently and completed in any order. Sequencing operations requires multiple rounds. Within ExaMPI, these sub-requests are represented by specialized internal requests that enable them to have a reference back to their round, as well as give them their own set of methods that allow them to behave appropriately. The rounds themselves have no reference for where they reside in the schedule; they only know if they have a next round to launch when their time comes. If the round  does not have a subsequent round to launch, the progress engine assumes that it is the last round, and marks the schedule request as complete. This assumption comes from the basis that it is not possible to build a schedule with multiple final rounds. As per the flow diagram in Fig.~\ref{fig:schedule}, in order to make a new round, the rounds must be first linked. Only the last round is allowed to not have a link to another round. The design (and user-level API) forbids a round from having multiple next rounds, which in turn prevents multiple final rounds.Ideally, the user-built schedule would behave logically as if the user called the normal MPI functions in the same order as defined in the schedule.

\subsection{Progressing Operations}
\begin{figure}[t]
    \centering
    \includegraphics[width=0.8\linewidth]{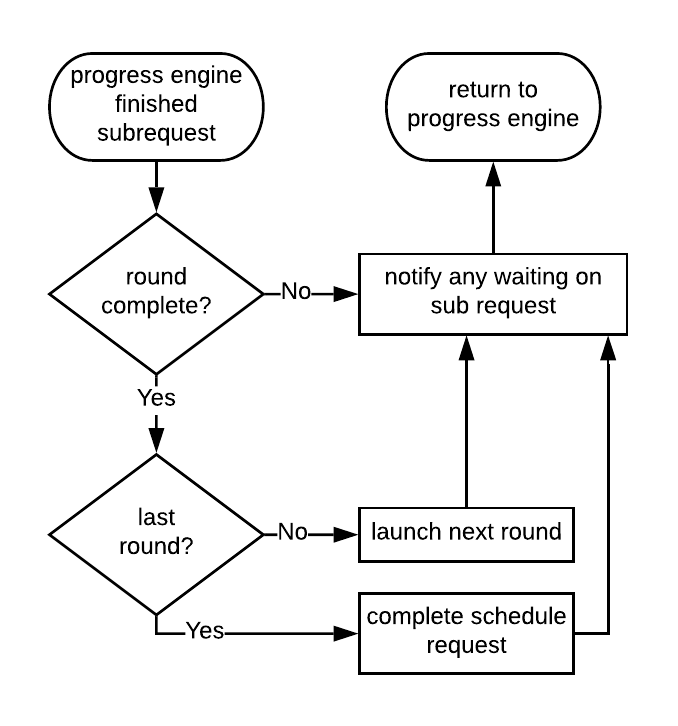}
    \caption{Progress engine progressing a request}
    \label{fig:release}
\end{figure}

Internally, the sub-requests added to the rounds are a special type of request that behave differently when the progress engine finishes with them. To the engine itself, it is not aware that there is such a distinction. To make the sub-requests unique, they override the function that is called when the progress engine wants to release a request it has finished. Instead of only marking itself as complete and notifying anyone waiting on this request, these special requests also interact with their respective round. When a sub-request is completed, it tells its round that it is done, which increments a completion counter inside the round. To avoid a race condition from a multi-threaded progress engine, only one sub-request may access the round's counter at a time. After successfully incrementing the counter, the sub-request checks with the round to see if it is the last one to arrive. If it was, the sub-request then tells the round to progress to the next round. The current round will then take care of launching the next round, which means launching its batch of sub-requests\footnote{While operations in a round should not have a specific ordering, the launching of a round will likely result in a first-in, first-out style of starting for the sub-requests. Additionally, the completion order will be purely up to how to the progress engine progresses the requests.}. But, if the round does not have a next round, it assumes that it is the final round (as mentioned above) and that it is the round responsible to marking the original schedule request as completed. Fig.~\ref{fig:release} shows the aforementioned flow from the progress engine's side.

\section{Results}
\label{sec:results}

\begin{figure}
\centering
  \includegraphics[width=\linewidth]{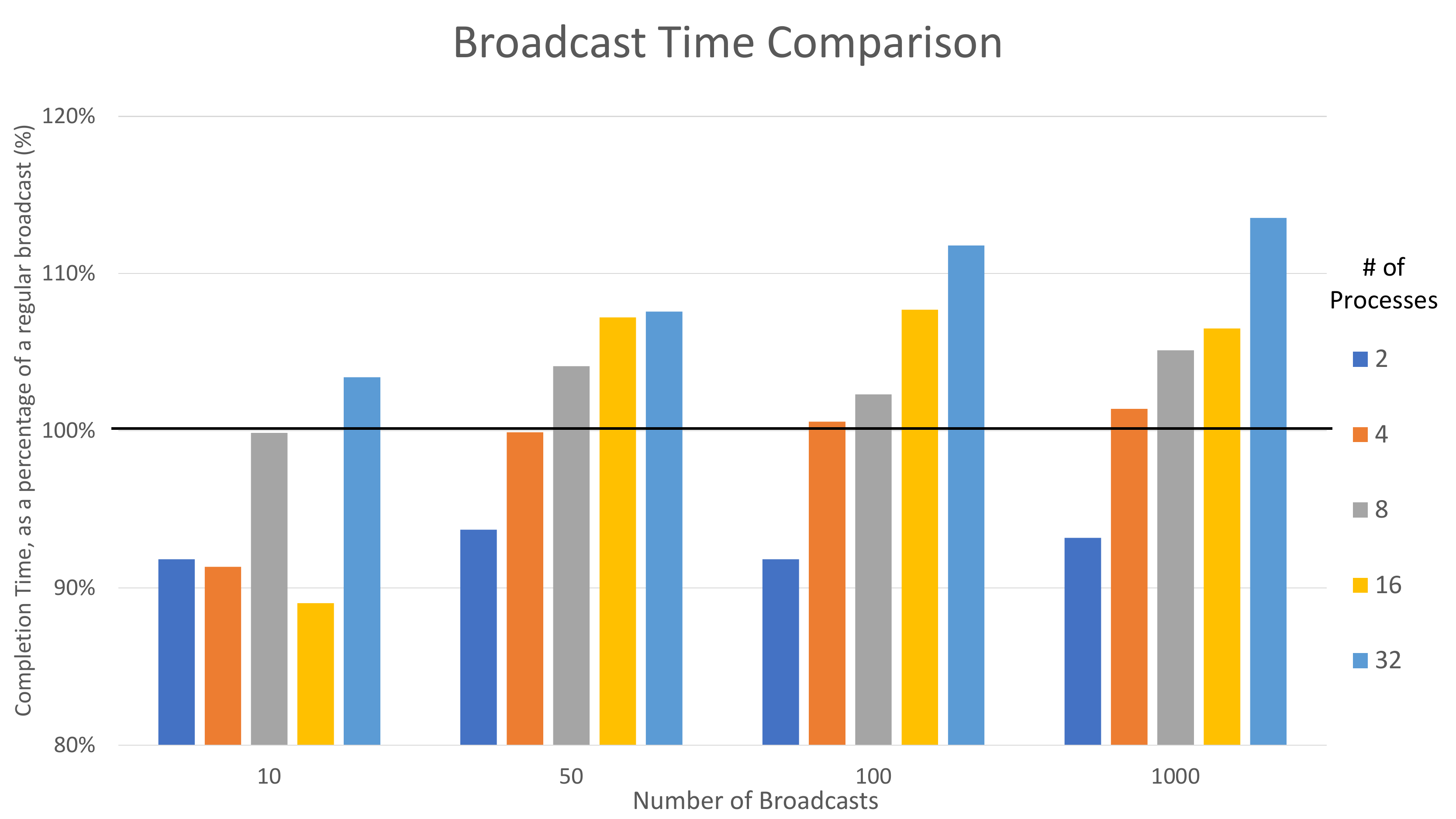}
  \caption{A comparison of regular broadcast times versus schedule broadcast times. The black line represents the time of a regular broadcast. Any times over the line are slower than the regular broadcast and any times under the line are faster.}
  \label{fig:results2}
\end{figure}

We have evaluated our implementation using ExaMPI~\cite{CARLA2019}. We first measured the overhead of schedule creation and our measurements indicate that the schedule creation overhead is  low (less than 1 ms, on average). Fig.~\ref{fig:results2} shows time required to complete  the standard blocking broadcast compared to our broadcast with schedule. We ran multiple tests with varying number of broadcast (10, 50, 100, 1000) to understand the benefits of using a schedule multiple times and with a varying number of processes (2, 4, 8, 16, 32) to gain some insight into scalability. In this figure, a bar of 150\% means that using a scheduled broadcast in this scenario, on average, was 50\% slower than a regular MPI broadcast; a bar of 50\% means that using a scheduled broadcast is 50\% faster than a regular MPI broadcast.

From the figure, we see that these early results indicate that as one uses the schedule more often, there is an insignificant increase in delay, which means that the user is not increasing time to completion when using a schedule. While increasing the number of processes does appear to increase the time of executing a scheduled broadcast, this slight extra time is likely the result of  the small overheads of the schedule itself.

\begin{figure}[t]
  \centering
  \includegraphics[width=\linewidth]{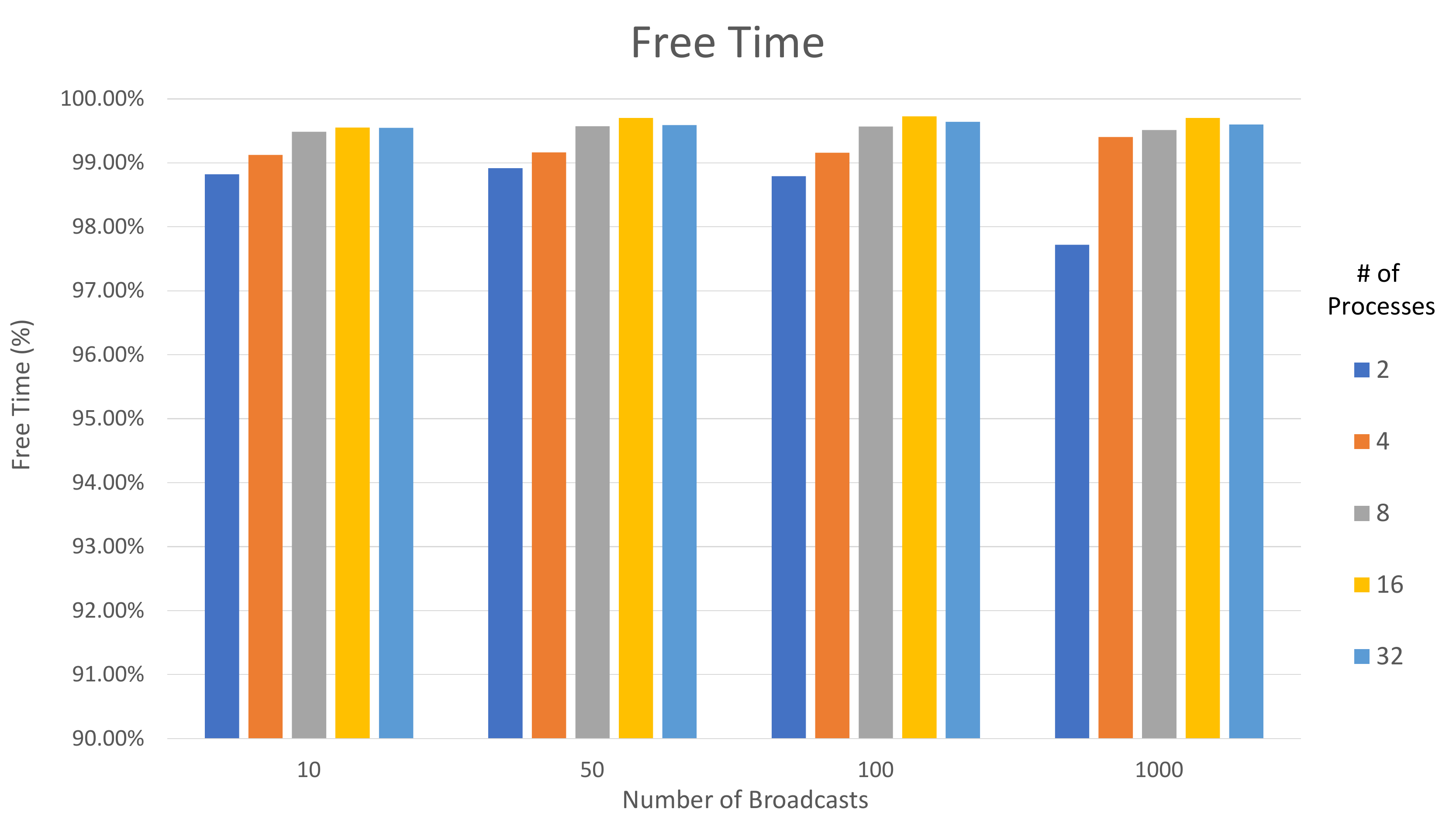}
  \caption{Average free time when using schedule (higher is better)}
  \label{fig:results1}
\end{figure}

With ExaMPI's strong progress engine, we were successfully able to offload the schedule to the progress engine and achieve a 98\%+ gap to do computations, as shown in Fig.~\ref{fig:results1} This is important because it means that as we starting scaling schedules to even more processes, we have that much more time to overlap computation with the communications done in the schedule. Having that  free time means that users can achieve increased overlap of communication and computation with  schedules.

Interestingly, it appears that in the case of two processes (where a broadcast is only one send and one receive), the scheduled broadcast was faster, perhaps indicating an extra overhead in ExaMPI; we will revisit that performance differential by further study of ExaMPI.

\section{Extensions and Future Work}
\label{sec:future}

MPI supports no asynchronous notification methodology, such as the event-driven model shown in PERUSE \cite{PERUSE} and MPI/RT (a priority, event model) \cite{DBLP:journals/concurrency/SkjellumKDWPCHHCR04} efficiently. While PERUSE was designed for event-driven notification of MPI implementation information and worked successfully with certain MPIs such as MPI/Pro \cite{mpipro} and Open MPI, it was never adopted by the Forum.  Instead, the Forum has encouraged polling-base notification of events for MPI-4 tools \cite{TOOLSCALLBACKSMPI4,MPIISCBOF2019} (slides 18--27).  MPI could well be elaborated to support event-driven programming, using the thread-levels described in the MPI-4 tools proposal for callbacks (with extensions to allow hardware offload through downcalls vs.\@ upcalls).  Such callbacks would allow events generated by MPI applications to be local, groupwise, or external in origin.  For instance, this model allows layering on top of MPI of certain operations, such as MPIX\_COMM\_REVOKE\footnote{Despite its mention here, this paper does not endorse MPIX\_COMM\_REVOKE as a good approach to error propagation in a fault-tolerant MPI.}  in ULFM \cite{DBLP:journals/computing/BlandBHHBD13},  to be implemented as a groupwise event that calls the error handler on a communicator across a group, rather than as a monolithic new API to be standardized.  Another example is the MPIX\_PREADY API proposed with partitioned communication \cite{DBLP:conf/supercomputer/GrantDLBS19,MPIISCBOF2019} (slides 38--45); it constitutes an event being raised by an MPI application to tell MPI that a message partition is complete and available to MPI for transmission.

The concepts presented here can be extended, with care, to MPI one-sided communication schedules as well, but require non-blocking constructors and destructors for windows and persistent versions of operations such as MPI\_WIN\_FENCE.  Some of us and others are studying such extensions to MPI.

A further opportunity for schedules defined using this API is the potential for offloading them completely to FPGAs.  Since schedules are persistent, their MPIX\_SCHEDULE\_COMMIT operation could include programmatic transfer to an FPGA, with the potential for either using a state machine on this device (or potentially compiling and loading new FPGA logic through partial reprogramming for such as schedule).

Finally, there is a desire to eventually make user-level schedules Turing complete. While the authors acknowledge that the design presented in this paper represent a DAG more than a Turing complete implementation, the user-level schedules are only a few steps from Turing complete. In the future, we hope to implement the ability to repeat operations in a schedule for a user-specified number of times and provide conditionals to decide whether a certain operation will be performed or not.  

\section{Conclusion}
\label{sec:conclusion}
The introduction of persistent collective communication and forthcoming partitioned point-to-point communication (which provides real channels for MPI) has opened the opportunity to introduce another important extension to MPI; namely, user-level scheduled communication.  Generalized requests, introduced in MPI-2, do not meet the needs for effective generalized communication extensions.  

This paper introduces user-level scheduled communication, an extensible interface motivated in part by internal concurrent schedule interfaces in certain MPI libraries. Background, motivation, design, implementation and qualitative performance modeling were described.
Implementation with the ExaMPI research MPI implementation is highlighted.
The value of strong progress in supporting these operations is mentioned.

The schedules described here provide DAG-type operations including the ability to do multiple rounds of communication, and to include, hierarchically, both point-to-point and collective operations inside schedules.  This differs from and extends the schedules implemented in libraries such as Open MPI, which are point-to-point based, and not part of the user API.  Benefits of this extension to MPI include the ability for high-performance collective operations to be added without working within an MPI library itself, as well as introducing new communication patterns at high performance.  These operations would also be cross-MPI portable if these extensions become standardized in future.  

Early performance results indicate that schedule formation is not overly expensive, considering that schedules will normally be used many times once created.  Also, performance indications show that schedules work well with ExaMPI's strong progress engine, providing the potential for overlap of communication and computation during the time schedules are active.

While this paper confines itself to persistent message passing, concepts described here  generalize to support schedules of  non-persistent MPI operations, as well as  one-sided communication operation. One-sided operations can be scheduled provided certain additional operations are added to the standard (e.g., fully non-blocking window creation).
Further, if combined with
event-based extensions to MPI (asynchronous notification), these operations provide a basis for full
task-based parallel programming with message passing and will help enhance scheduling and triggering of MPI+X operation sequences (e.g., MPI+CUDA).


\section*{Acknowledgment}
This work was 
supported in part by the National Science
Foundation under Grants Nos.~CCF-1925603, CCF-1822191, and CCF-1821431. Any opinions, findings, and
conclusions or recommendations expressed in this material are those of
the authors and do not necessarily reflect the views of the National
Science Foundation.

This work was part-funded by the European Union's Horizon 2020 Research and Innovation programme under Grant Agreement 801039 (the EPiGRAM-HS project).

The authors    wish   to thank Dr.\@ Puri Bangalore for his helpful feedback on this paper.



%

\nocite{skjellum1994extending,Geist:1996:MEM,SkjellumPDPTA98}
\bibliographystyle{IEEEtran}
\bibliography{references.bib}





\end{document}